# On Aperture-Friction Networks

Ghaffari, H.O.
*Department of Civil Engineering and Lassonde Institute, University of Toronto, Toronto, Canada*

**ABSTRACT:** The shear strength and stick-slip behavior of a rough rock joint are analyzed using the complex network approach. We develop a network approach on correlation patterns of void spaces of an evolvable rough fracture (discrete time contact areas-crack type II). The several properties of aperture-friction networks are presented. The mechanical and hydro-mechanical behaviors of the rock interfaces are probed through the corresponding aperture-friction networks. Next, a model based on self-organizing nature of the surfaces is represented to capture the evolution of friction networks. Also, the curvatures of displacement fields are determined and critical curvature profiles are connected through links to form ridge-networks. To generate the ridge-networks, we consider the interactions of pairs though a directed network framework, namely, dilatancy (divergence) and deviotric components (vorticity) of local maxima of curvature profiles. The correlation of the characteristics of generated networks with synthetic acoustic signatures of frictional interface is postulated while the interactions of critical pairs before and after frictional sliding highlighted.

## 1. INTRODUCTION

Evolution of macroscopic friction in frictional interfaces originates from the sequence of contact area variations [1-6]. The formation and rupturing of new contact areas (i.e., bonds, junctions, asperities) forming between two surfaces results in stick-slip motion. Classical characteristics of stick-slip motion are fast frictional strength drops and energy released as spikes in ultrasonic wave forms [6]. Recent findings also suggest that stick-slip motion is related to collective interactions of contact areas [5, 7-11]. Hence, sheared systems-particularly in the form of granular elements show an anisotropic and long correlation around shear cracks [11-14] .One way to characterize the contact patterns and their collective behavior is to discretise the patterns into separated "contact or non-contact" patches. The strings or patches, as the hypothetical objects, are the profiles constructed by collecting elements of the system (pure contacts or relative contact areas) along a thin line (i.e., a ribbon). Considering only pure contact cells represents stair-like patches, which we called contact patches or contact strings. "Patches" are characterized by mapping the interactions onto networks, in which each patch is mapped onto a node and the corresponding interactions, connecting the nodes, are the links. The aforementioned technique has been used successfully in analysis aperture of a rock joint. It has been shown that the unraveled characteristics of the friction networks scales with the mechanical and hydro-mechanical features of the interface [10-12]. In this paper, we present more features of aperture-friction networks, among them modularity, assortatitvity and loops are of interest. Also, we look at the source mechanisms of critical points in displacement field and their possible pair-wise interactions. In the latter part we employ an idea of "dipole" interactions under the network's framework. Interestingly, we find loops in the friction networks scale with the node's degree. This is the main characteristic of friction networks where we found in over several friction patterns [15]. As another observation, the evolution of node's degree can be simulated under the Turing patterns as non-linear self-organizing systems. We recognized with-in module index of friction networks follow a spatial-periodic patterns, indicating Soliton wave solution [16] and probably related to the characteristics length of the interface. Outstandingly, the curvatures of displacements fields follow a gamma distribution which is compatible with the extended Omari's law [17]. We also show how interactions of sources scale with maximum acoustic activity of the interface.

## 2. THE NECESSITY OF THE STUDY

We discuss about the necessity of study of contact patterns (or aperture, asperity patterns). A simple mathematical analysis is represented with respect to contact mechanics and asperities interactions (for details see [18]). Let us assume two cases where two similar circular-contact areas (with radius a) formed between a rigid top surface and a linear half space with shear Modulus $G$ and Poisson's ratio $\nu$ (~relative contact areas are same; this is equal somehow to observed aperture). The mechanical interactions between two asperities are approximated by Cerruti solution (we assume the contacts are under lateral shear force):

$$\begin{bmatrix} A^1 & 0 & B_x^1 & C^1 \\ 0 & A^1 & C^1 & B_y^1 \\ B_x^1 & C^1 & A^1 & 0 \\ C^1 & B_y^1 & 0 & A^1 \end{bmatrix} \begin{bmatrix} F_{x1}^1 \\ F_{y1}^1 \\ F_{x2}^1 \\ F_{y2}^1 \end{bmatrix} = \begin{bmatrix} \delta_{x1}^1 \\ \delta_{y1}^1 \\ \delta_{x2}^1 \\ \delta_{y2}^1 \end{bmatrix} \rightarrow \mathbf{MF} = \boldsymbol{\delta} \qquad (1)$$

in which,

$$r_1 = \sqrt{(x_2^1 - x_1^1)^2 + (y_2^1 - y_1^1)^2}, A = \frac{1}{8a}\frac{2-\nu}{G},$$

$$B_x^1 = \frac{1}{4\pi G}((2-2\nu)\frac{1}{r_1} + 2\nu\frac{(x_2^1 - x_1^1)^2}{r_1^3}$$

$$B_y^1 = \frac{1}{4\pi G}((2-2\nu)\frac{1}{r_1} + 2\nu\frac{(y_2^1 - y_1^1)^2}{r_1^3} \qquad (2)$$

$$C^1 = \frac{1}{4\pi G}(2\nu\frac{(y_2^1 - y_1^1)(x_2^1 - x_1^1)}{r_1^3})$$

The second case is similar to the first case (see Figure 1 b), however, a small difference in the position of the second asperity is induced. Further assumption in $x_1^1 = x_1^2 = y_1^1 \equiv 0; x_2^1 = x_2^2 \equiv 0$ where subscript and superscripts are local position index and case number (1 or 2), respectively. For a certain displacement and both case, shear force distribution among contact areas can be solved. If we assume, furthermore, $y_2^1 < y_2^2 \rightarrow r_2 > r_1$ then $A_1 = A_2; r_2 > r_1 \rightarrow B_x^1 > B_x^2, B_y^1 > B_y^2$. The later inequality implies the final shear stress distribution is different for each case as if we only change the pattern of contacts. For complex contact patterns where we have different contacts areas and connectivity of zones, M is so complex and analytically cannot be inferred. Then, study of friction is in a direct relationship with study of contact patterns rather than pure magnitude of contact areas. With a simple transformation of the surface into a network (graph) frame, we see the connectivity patterns of the network are dramatically change as well as other characteristics of the networks. In addition to this, we distinguish the observations of long-range correlations in sheared – granular systems which have been approved numerically and experimentally [11-14]. Long-range correlations in the fluctuations of shear strain induce the "excitation of additional elastic modes". The strong correlation in the direction of shear lowers the effective resistance to rupture in the direction of shear. We characterized this correlation with networks.

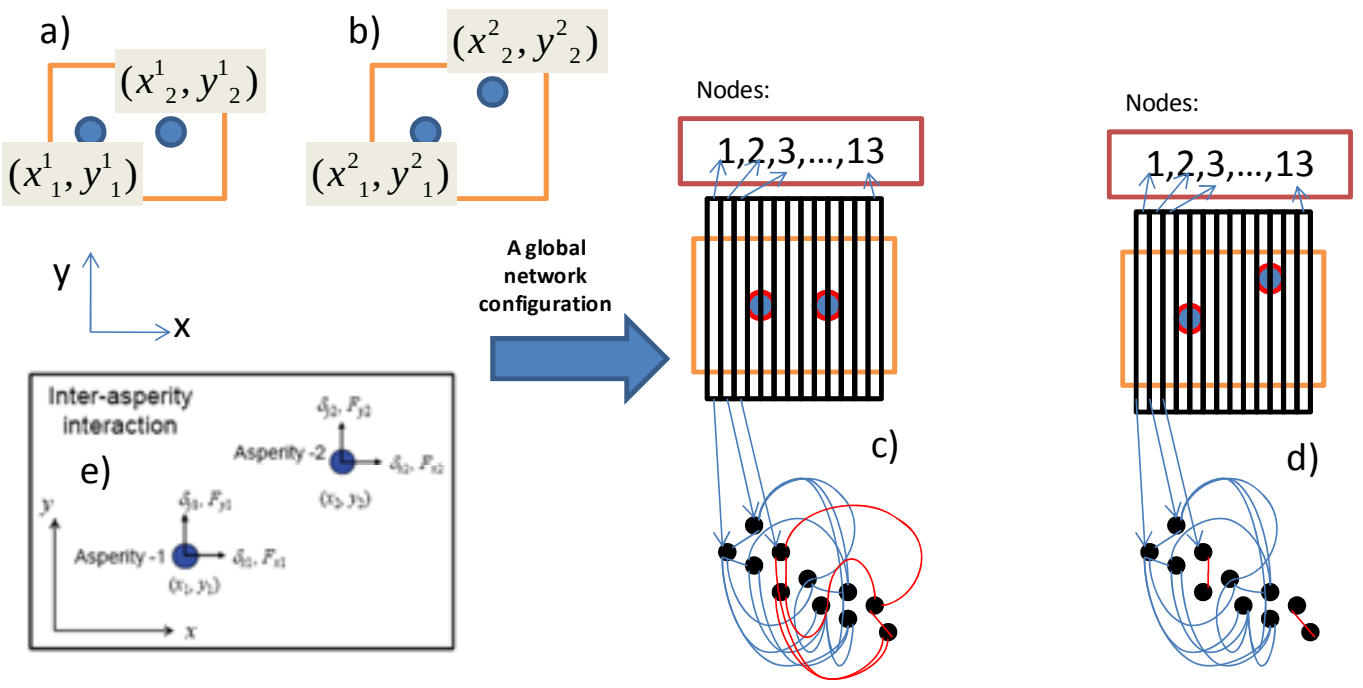

Fig 1. a,b) two similar circular contact areas with radius rand with different distance set up in a rigid plate.; c,d) transforming profiles (black rectangulars) to nodes and connecting them based on similarity of profiles(~ribbons) ;e) the components of forces and displacements due to mechanical interaction among contact areas.

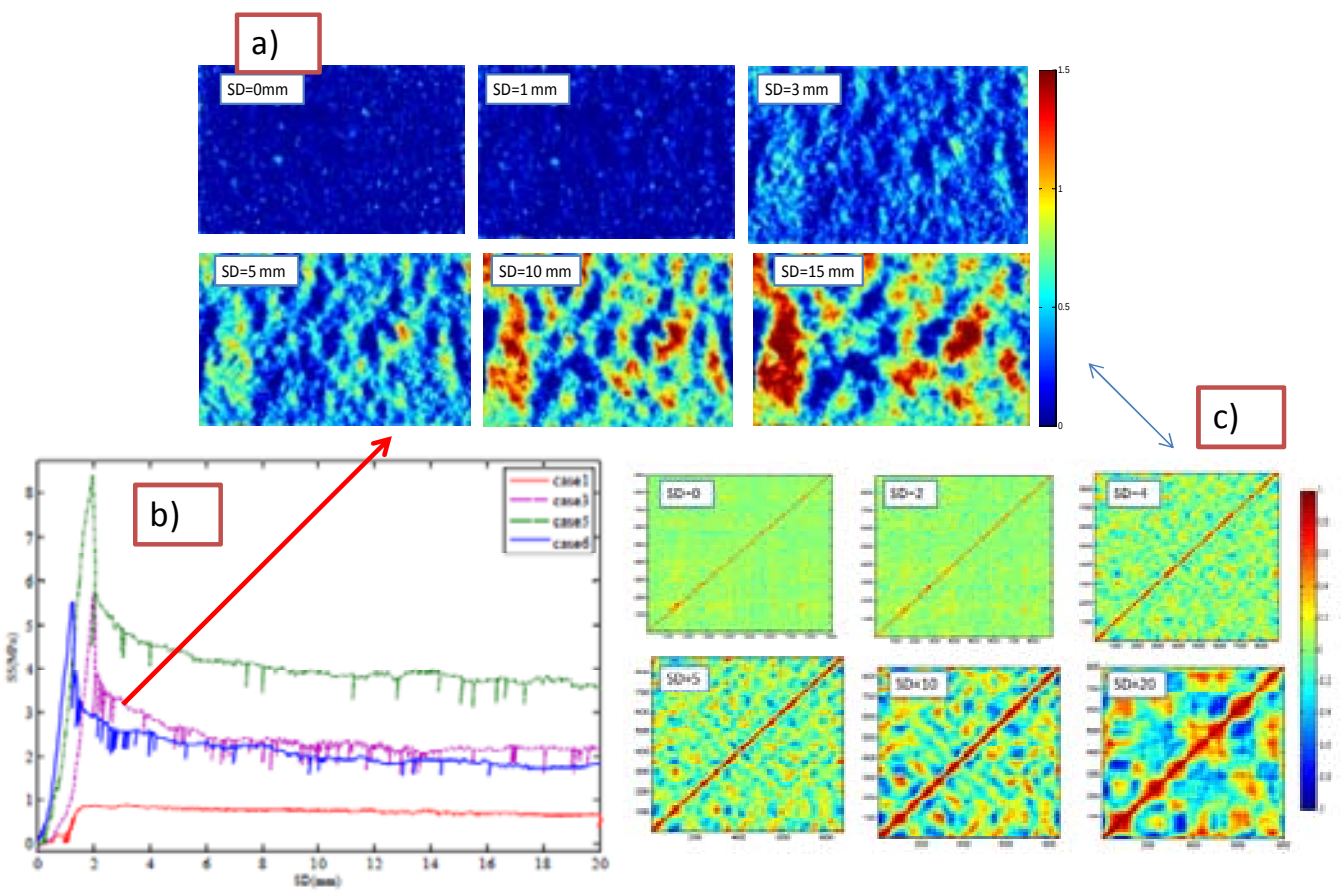

Fig 2. a)aperture patterns of a rock interface under 3Mpa normal load ;b) shear strength developments versus shear displacements (SD in mm) [19,20] ;c) correlation patterns of aperture patches in perpendicular aperture profiles.

## 3. DATA

The laboratory test procedure involved preparing a rough fracture, measuring the morphology of halves with a laser scanner, and measuring permeability. The rock used was granite with a unit weight of 25.9 kN/m3 and a uniaxial compressive strength of 172 MPa. An artificial rock joint was made by splitting the specimen mid-height with a special joint-creating apparatus [19, 20]. The sides of the joint were cut down after it was created. The final sizes of the samples were 180 mm in length, 100 mm in width, and 80 mm in height. A virtual mesh with a square element size of 0.2 mm was spread on each surface, and the height at each position was measured with a laser scanner. In Figure 2, we have shown the evolution of aperture patterns and also the developments of the related correlation patterns.

## 4. APERTURE-FRICTION NETWORKS

We use correlation metrics over 2D relative contact patterns (figure 2a). To set up a non-directed network over 2D contact areas in a certain time step, we considered each patch of measured contact areas perpendicular to shear direction as a node. Each profile has *N* pixels where each pixel shows the relative-contact area of that cell. Then, we define correlations in the profiles by using:

$$C_{ij} = \frac{\sum_{l=1}^{N}[A_i(l) - \langle A_i \rangle].[A_j(l) - \langle A_j \rangle]}{\sqrt{\sum_{l=1}^{N}[A_i(l) - \langle A_i \rangle]^2}.\sqrt{\sum_{l=1}^{N}[A_j(l) - \langle A_j \rangle]^2}}, \quad (3)$$

where $A_i(l)$ is $i^{\text{th}}$ profile with $1 \le l \le N$. We may see each profile as a separated cycle of a spatial series (i.e., collection of cycles in *x*-direction). To map the obtained series, we define each patch as node. To make an edge between two nodes, relative-high correlated profiles are connected ($C_{ij} \ge r_c$) with non-direct links. To choose the optimum value of $r_c$, we notice that the aim is to reach or keep the most stable structures in the total topology of the constructed networks. Different approaches have been used such as density of links, the dominant correlation among nodes, motif density or correlation dimension. To choose $r_c$, we use a nearly stable region in rate of betweenness centrality (B.C) - $r_c$ space which is in analogy with minimum value in the rate of edges density. The later method has been used successfully in analysis of time-series patterns in network space [21]. We notice finding a nearly stable region in *B.C*- $r_c$ space satisfies dominant structures of contact patterns.

To proceed, we use several characteristics of networks. Each node is characterized by its degree $k_i$ (number of links connected to that node) and the clustering coefficient. Clustering coefficient (as a fraction of triangles (3 point loops/cycles) is $C_i$ defined as $C_i = \frac{2T_i}{k_i(k_i - 1)}$ where $T_i$ is the number of links among the neighbors of node *i*. Then, a node with *k* links participates on *T(k)* triangles. With respect to clusters, groups or communities in networks, we interest in detecting modules and their possible role in conducting frictional interfaces. Furthermore, based on the role of a node in the modules of network, each node is assigned to its within-module degree (*Z*) and its participation coefficient (*P*). High values of *Z* indicate how well-connected node to other nodes in the same module and *P* is a measure to well-distribution of links of the node among different modules [22]. To determine modularity and partition of the nodes into modules, the modularity *M* (i.e., objective function) is defined as [22-23]:

$$M = \sum_{s=1}^{N_M}[\frac{l_s}{L} - \left(\frac{d_s}{2L}\right)^2], \quad (4)$$

in which $N_M$ is the number of modules (clusters) , $L = \frac{1}{2}\sum_i k_i$ , $l_s$ is the number of links in module and $d_s = \sum_i k_i^s$ (the sum of nodes degrees in module *s*) . With using an optimization algorithm (here we use Louvain algorithm), the cluster with maximum modularity is detected. To describe the correlation of a node with the degree of neighboring nodes, assortative mixing index is used:

$$r_k = \frac{\langle j_l k_l \rangle - \langle k_l \rangle^2}{\langle k_l^2 \rangle - \langle k_l \rangle^2} \quad (5)$$

where it shows the Pearson correlation coefficient between degrees $(j_l, k_l)$ and $\langle \bullet \rangle$ denotes average over the number of links in the network.

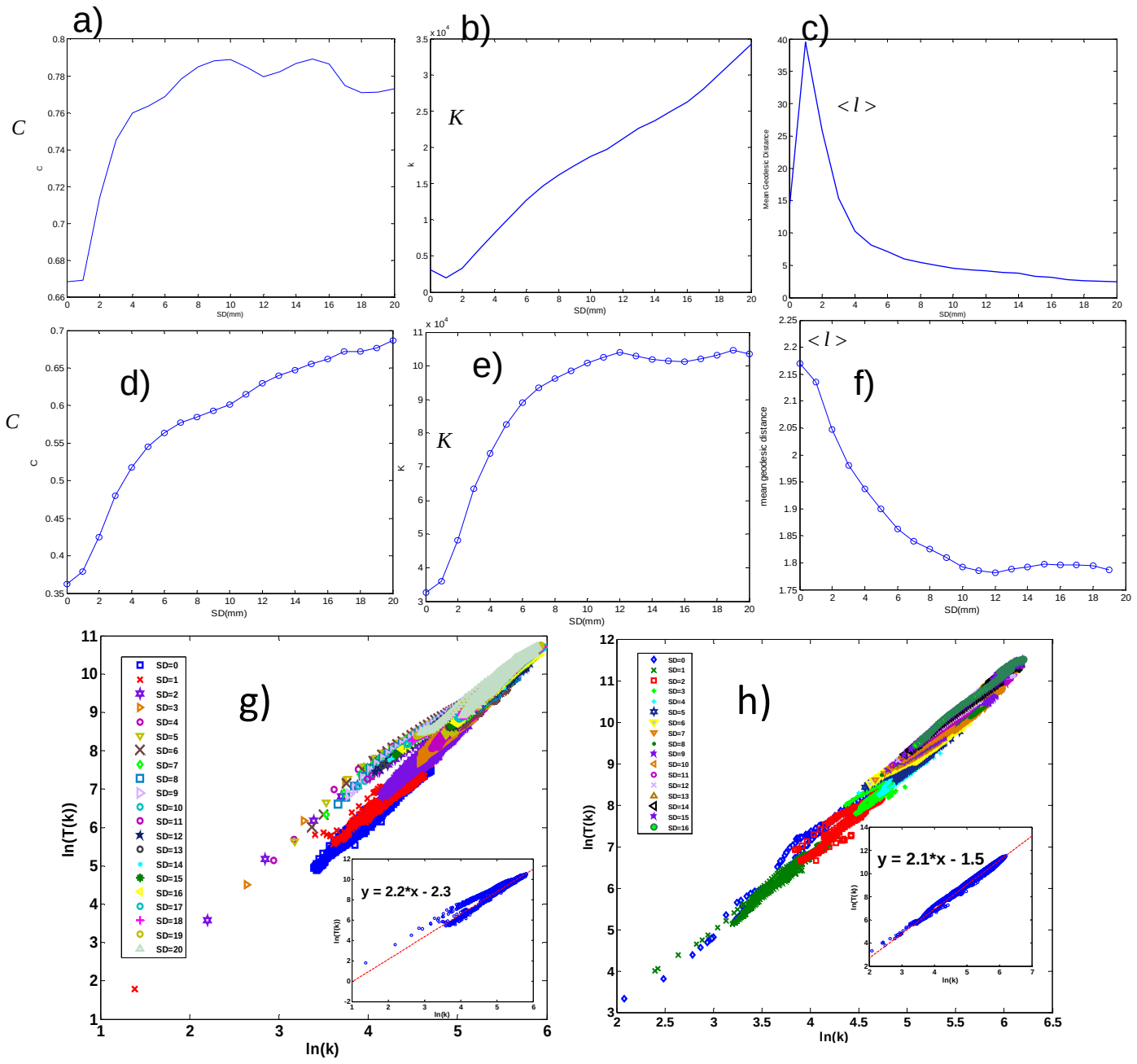

Figure 3. Characteristics of the friction-networks in parallel profiles-networks to shear direction (a -c) :a) Clustering coefficient-Shear Displacement (SD in mm);b) Number of edges-SD ; c) Average path length-SD; in transverse profiles-networks to shear direction (d -f) ; (g) X-profiles :scaling of triangles (T(k)-i.e., loops) with node's degree(k) as a power law with $\beta \approx 2.2$ (inset shows the best fit linear line to collapsed Data set in natural logarithmic scale) (h) Y-profiles (parallel to shear) a nearly same scaling with $\beta \approx 2.1$.

In Figure 3, we plotted the three parameters of the networks through 20 mm shear displacements. After a transition stage, the formation of 3-node loops in a parallel direction reaches a quasi-stable state, while the perpendicular networks follow a growing trend (Fig. 2a). This trend is reversed in the evolution of the nodes' degree (Fig. 3b), which shows the growth of the profiles' long-range correlations. The characteristic length of networks exhibits a rapid drop after passing the interlocking step, where the asperities are locked up. This stage occurs just after the peak point of shear stress-displacement. However, for the perpendicular-direction case analyzed, the transition point at the same stage displayed a softer change (Fig. 3c). The similarities among the three presented characteristics of the networks occur in the transformation from 1 mm to 2 mm of displacement, where the rock joint under a certain value of normal stress passes the maximum frictional strength (see Figure 2b). Furthermore, a consideration of 3-point cycles (*T*-triangle loops) versus the nodes' degree shows a power low scaling (Fig. 3g, h):

$$T(k) \sim k^{\beta}, \tag{6}$$

where the best fit for the collapsed data set reads $\beta \approx 2 \pm .3$ (which we call a coupling coefficient of local and global structures). With some mathematical analysis (see appendix), one can show that adding *m* edges increases the number of loops with $\beta^2 m^{\beta}$, which indicates a very congested structure of global and local sub-graphs during shear rupture. Also, we notice $C(k) \sim 2k^{\beta-2}$, so that for $\beta < 2$, a possible hierarchical structure can be predicted .

Next, we focus on the highlighted patterns and obviously our results can be investigated to other patterns (presented patterns in figure 2). Obviously, one can follow a self-organized correlation patterns which can be analyzed in terms of nonlinear dynamics methods. One of the methods to describe the self-organization nature of patterns is using a series of coupled equations in terms of inhibitor-activator (Turing patterns). Then, we can describe the evolution of number of edges (or node's degree) with respect to Turing patterns, which it reads:

$$\begin{aligned} \frac{\partial}{\partial t} k_i(t) &= f(k_i, u_i) + \varepsilon_k \nabla^2 k(x_i, y_i, t) \\ \frac{\partial}{\partial t} u_i(t) &= g(k_i, u_i) + \varepsilon_u \nabla^2 u(x_i, y_i, t) \\ f(k_i, u_i) &= \gamma(a - k_i + k_i^2 u_i) \\ g(k_i, u_i) &= \gamma(b - k_i^2 u_i) \end{aligned} \tag{7}$$

in which $k_i$, $u_i$, $\varepsilon_k$ and $\varepsilon_u$ are node's degree ,a system attribute (as well as displacement or temperature) mobility of patterns coefficients ,respectively. With $\varepsilon_k$=1 and $\varepsilon_u$>1, we obtain Turing patterns [24].The observed Turing patterns in the evolution of friction-networks is a link to the large scale observed self-organized criticality in earthquakes, as well as Omari's law. Since the 1980s, the theory of self-organized criticality (SOC) over the time-series of earthquakes (either in laboratory scale or large scale) has been suggested that plays a significant role in diverse avalanche-like (or "crackling noise") [25]. In our case, the interface is evolved in a way that with the minimum variation of contact areas reaches to the optimum quasi-stable regime (the best possible robust contact patterns).

Amazingly, plotting the evolution of within module index in *x*-*t*-Z(*x,t*) shows a periodic nature (Figure 4f). We obtained the periodic property of Z*(x,t)* in other real-time contact areas [15]. We believe the periodicity of within-module degree and somehow participation coefficient -with respect to nearly the invariant nature of spatial-periodicity- are related to the characteristic length ($\lambda_c$) of the interface. In figure 4c, the evolution of the asssortativity indicates that turning point of the index corresponds to the slip weakening distance (in slip displacement-shear stress plot). Then, the shear rupture energy can be stated as: $G_c \sim \Delta\tau_b D_c \sim \Delta r_k D_c$ in which $\Delta\tau_b$ and $D_c$ are the breakdown stress drop and critical slip displacement. In real-time contact measures of PMMA‘s interfaces, the same role of the edges correlation ("hubnnes") can be pursued [15].

We notice that during the slip weakening distance, the hubs of aperture-friction networks attract other "poor" node such that the modularity of the systems increases. It is noteworthy, the spatial non-uniform growth of loops (T (k)) through the interface for X-profiles –figure 4a-represent the aggregation of loops around certain profiles. Notably, for Y-profiles the modularity scales with the interlocking step and the variation of permeability changes (figure 5c). In Y-profiles, generally speaking the periodic characteristics in modularity parameters are not observed; indicating the most response to mechanical stress is due to the perpendicular profiles. We also notice the spatial distribution of loops in Y-aperture friction networks are completely different from X-patches while general scaling with the node's degree is preserved (Eq.6). Our observations show approximating an interface over X-patches with 1D for understanding the rupture evolution is enough while for fluid-flow study, parallel with the shear direction, the Y-profile's approximation would be sufficient. Following figure 5a, It turns out the transition to the slip-weakening stage is scaled with lowering the loops in Y-aperture friction networks. The excellent collapsing of data set in *log(B.C)*- *C* parameter space (figure 4b and 5b)–either in X or Y profiles- shows a unique universality in local energy flow; *i.e.*, the interfaces evolve so as increasing fraction of triangle-loops lowers local information flow.

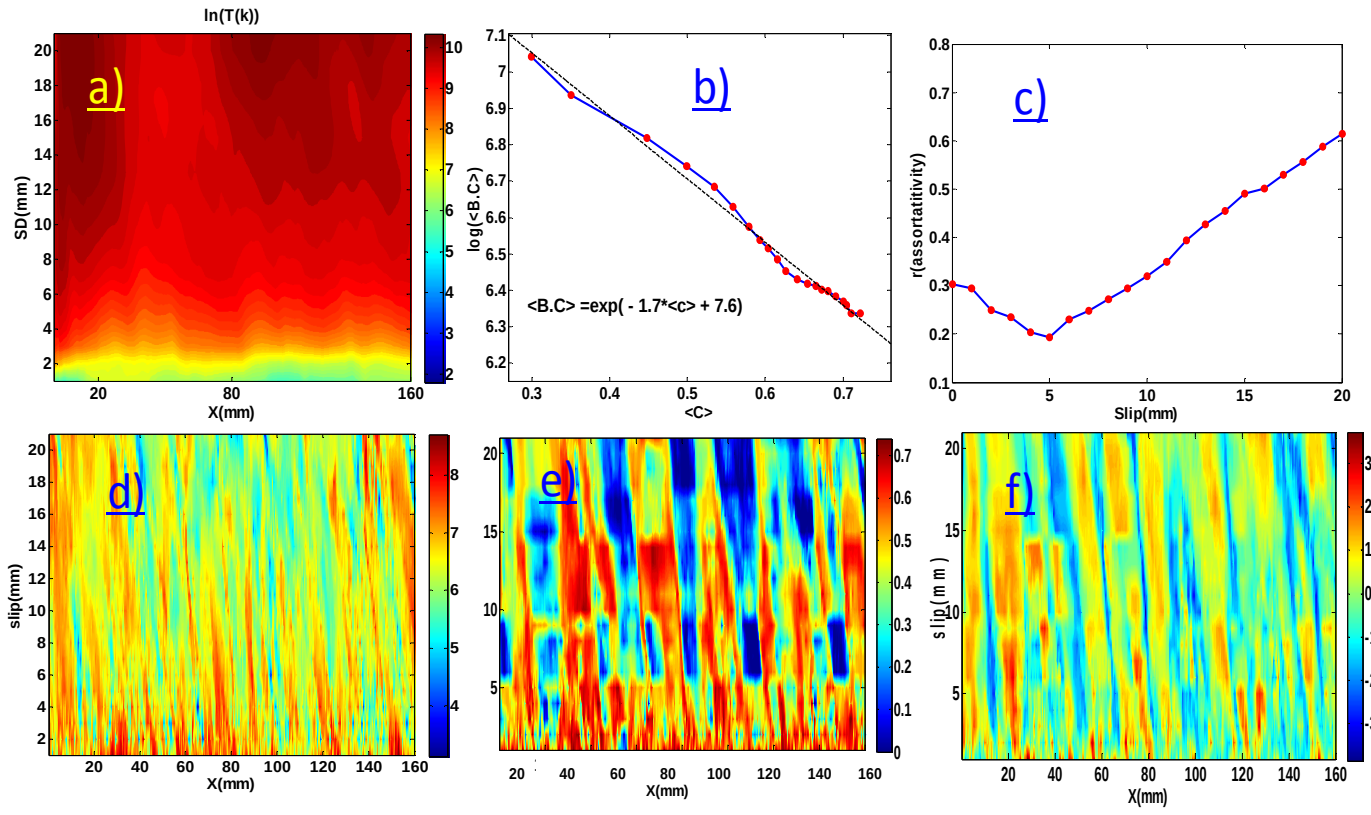

Fig 4.a) logarithm scale of loops variation with slip and along X;(b) scaling of betweennness centrality (as a local measure of information flow) with clustering coefficient ;(c) assortativity of networks with slip; It shows generally our networks are assortatitve ; Periodic patterns of the rupture surface: d) logarithmic scale of beteweness centrality (B.C) through perpendicular aperture profiles(the vertical axis is shear slip) ;e) participation coefficient (P) in slip-X space ;c) within module degree (Z) in slip-X space .

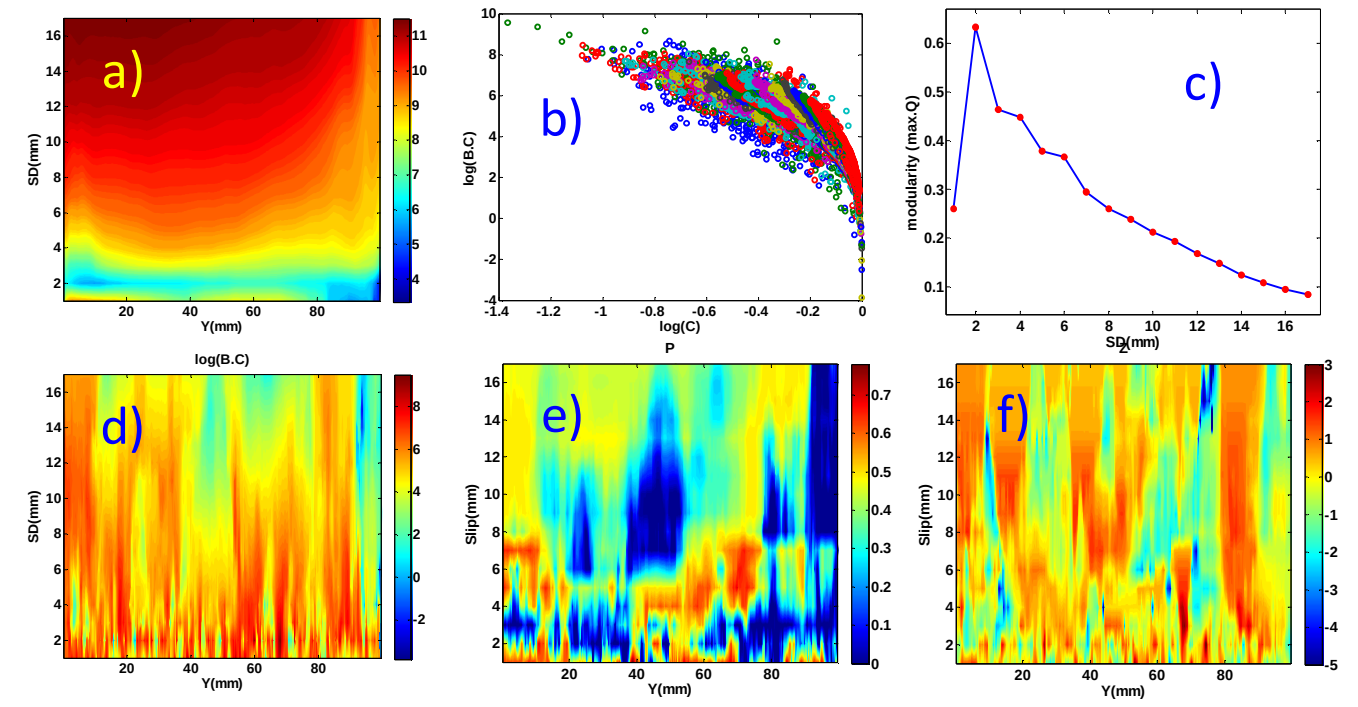

Fig 5.a) logarithm scale of loops variation with slip and along Y;(b) scaling of betweennness centrality (as a local measure of information flow) with clustering coefficient ;(c) maximum modularity (Q) with slip; d) logarithmic scale of beteweness centrality (B.C) through perpendicular aperture profiles (the vertical axis is shear slip); e) participation coefficient (P) in slip-Y space; c) within module degree (Z) in slip-Y space.

## 5. CURVATURE FIELDS AND PAIRWISE INTERACTIONS

As the last part of our analysis, we shed light the universality of curvature of deformations' components in shear strain (i.e., displacement). We notice the extermum curvatures of the displacements of particles are the points or zones of energy stagnations (or critical points). In fact, ridge networks or highly curved shells (or plates as well as crumpling phenomena) present a new field in modeling and investigation of materials [26].

With respect to frictional interfaces, shear displacement components is given by $\Delta \mathrm{u} = (\Delta x, 0, \Delta z)$ where we assumed interface does not have any shear component in y-direction (figure 6a). The curvature of any point in the displacement field is given by: $\kappa = \nabla^2(\Delta \mathrm{u})$. In Figure 6c, the distribution of curvatures for the first case study has been shown. The excellent data collapse implies that $\kappa$ does not depend on the speed of the shear rupture and the universal scaling function can be approximated by a gamma distribution: $P(\kappa) \sim \kappa^{-(1-A)} \exp(-\kappa/B)$ -$A$ and $B$ are the fixed parameters. The same relation is obtained for discrete-contact measurements. This relation is comparable with distribution of waiting times between earthquakes of energy larger than a threshold level [17]. Subsequently, one can assume the origination of generalized Omori's law is the result of critical points as the stagnation points of energy in energy landscape of the interface. Let's now consider another scenario where we detect critical points ("singular" or "stagnation") in the displacement field of a system under shear forces. We may deduce such a displacement field with $u = (\Delta x, \Delta y, \Delta z)$ where $\Delta x$ =constant with respect to quasi-static slip rate $\Delta y$ =0 and $\Delta z$ =aperture change (or contact area variation).

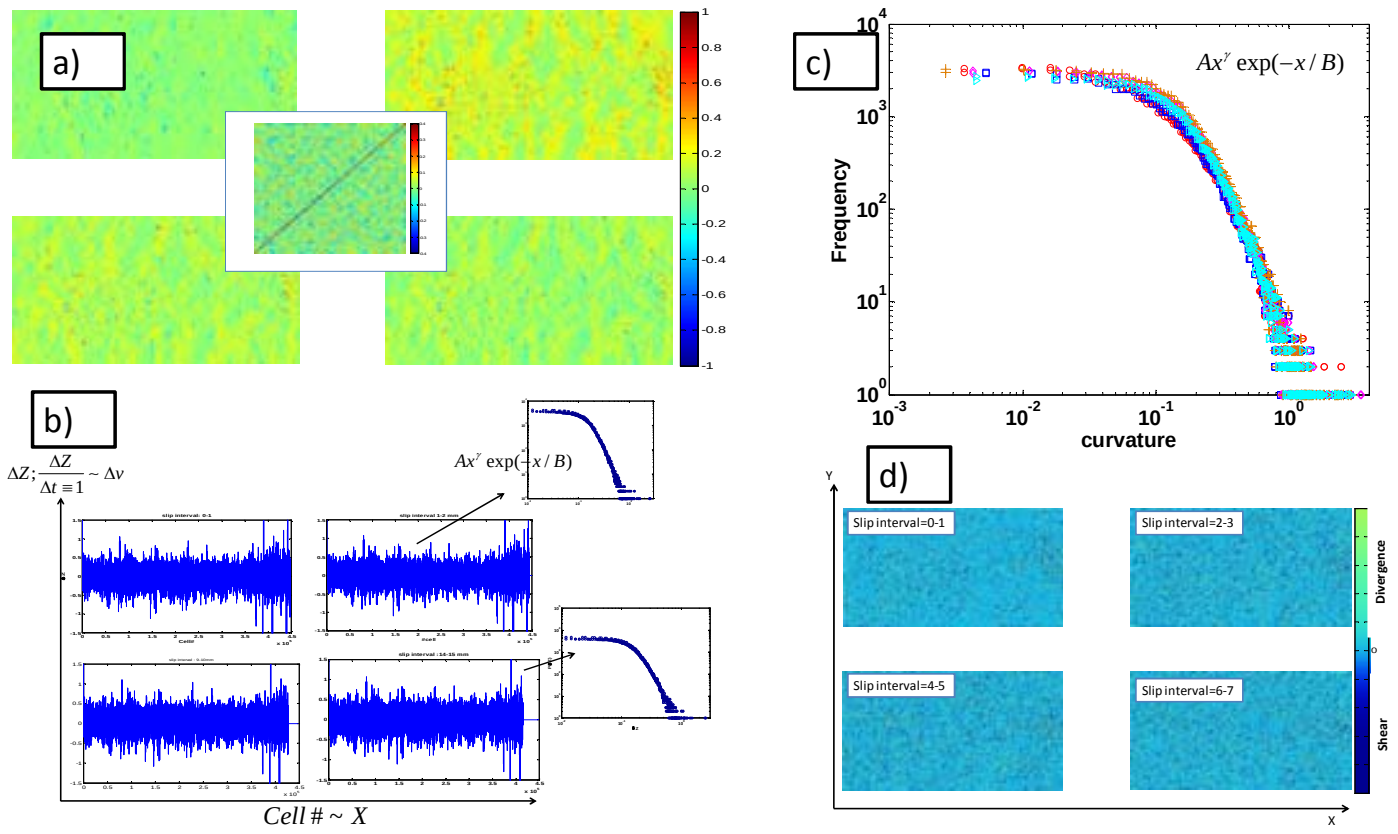

Fig 6. analysis of shear strain deformation (over successive displacements of interface) :a) aperture (deformation in $z$ direction) in 4 deformation statutes (the inset is correlation patterns of deformation patches in slip from 14 to 15 mm) ;b) wavelike configurations of deformation fields in 4 different shear deformation ;c)collapsing all curvatures of patches of strain in a Gama distribution (laplacian of deformation in z direction) ;d) discritization of deformation fields into pure shear and pure Divergence (volumetric) components (we also characterize these "sources" with a geometrical network approach).

The detection of critical points in displacement fields can be accomplished by introducing local maxima curvatures (Laplacian of the field: $\nabla^2 u$ ). By defining the displacement field's curl and divergence, we can infer information about possible critical points with respect to the features of those points (figure 6d) , which we determine with divergence and curl of the displacement field $\phi = \partial_x u_x + \partial_z u_z$ and $\omega = \partial_z u_x - \partial_x u_z$ ,respectively. The former gives information about local dilation, while the latter gives information about shear. Here, the features refer to the pure shear or the pure tensile attributes. Figure 6 shows the aforementioned procedure on the available data set from the first data set. In figure 6d, first the critical points which satisfy $\kappa \geq <\kappa>$ are picked up and then classified in pure dilation or pure shear (~double couple) states. To characterize the evolution of the categorized points, we shed light the “dipole” or coupled-interaction of sources. To do this, for each radius an influence (*r*) source is assigned so as the interactions with other with-in zone sources are characterized as graphs. In figure 7, we illustrated the evolution of the sources (~nodes) in terms of node’s degree for *r=8*. Interestingly, reaching to the residual state is governed by minimum numbers of the sources while the interactions among them are maximums. Maximum number of pair creation occurs after stress drop, while before reaching to maximum Coulomb threshold level, pair creation shows a growing trend. A more interesting result is the duration of some nodes in preserving the pairs, indicating the asperities position and the periodic nature of events. For this case study, with increasing the effective zone (r), a continuous transition from assortative to disassortative state occurs (around r=8)-see figure 8b. Around the same radius, dipole-curvature networks represent maximum modularity with neutral correlation of node’s degree (figure 8a). Surface density of 4-points sub-graphs shows a fast jump at transition from 2-3mm displacement, comparable with the transition to slip weakening (maximum activity of sources) - figure 8c. The later observation is independent of the influence radius, representing a nearly same characteristic for all source points.

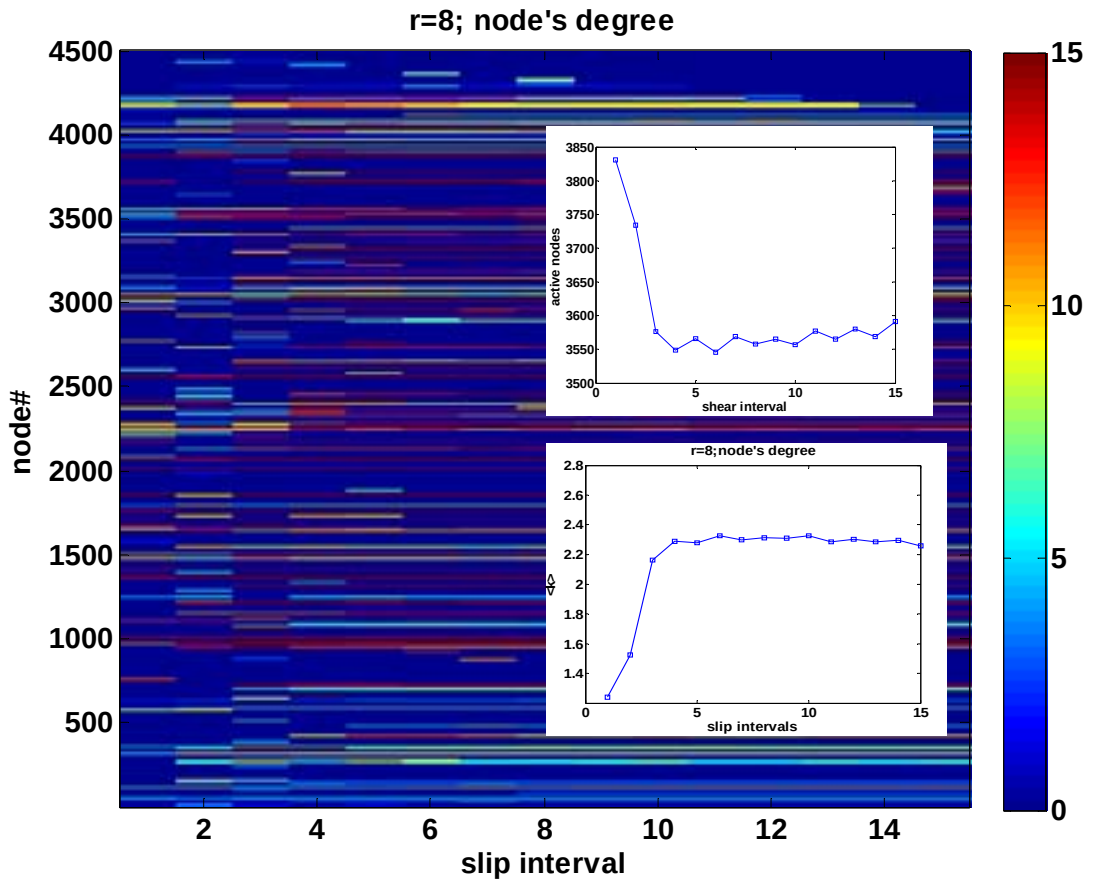

Fig 7. The results of tracking “pairs” over the reduced number of “events”. Here, we used an influence radius for each event (r=8 units). The creation or annihilation of interactions per each node is characterized by node’s degree (the number of links to each event).

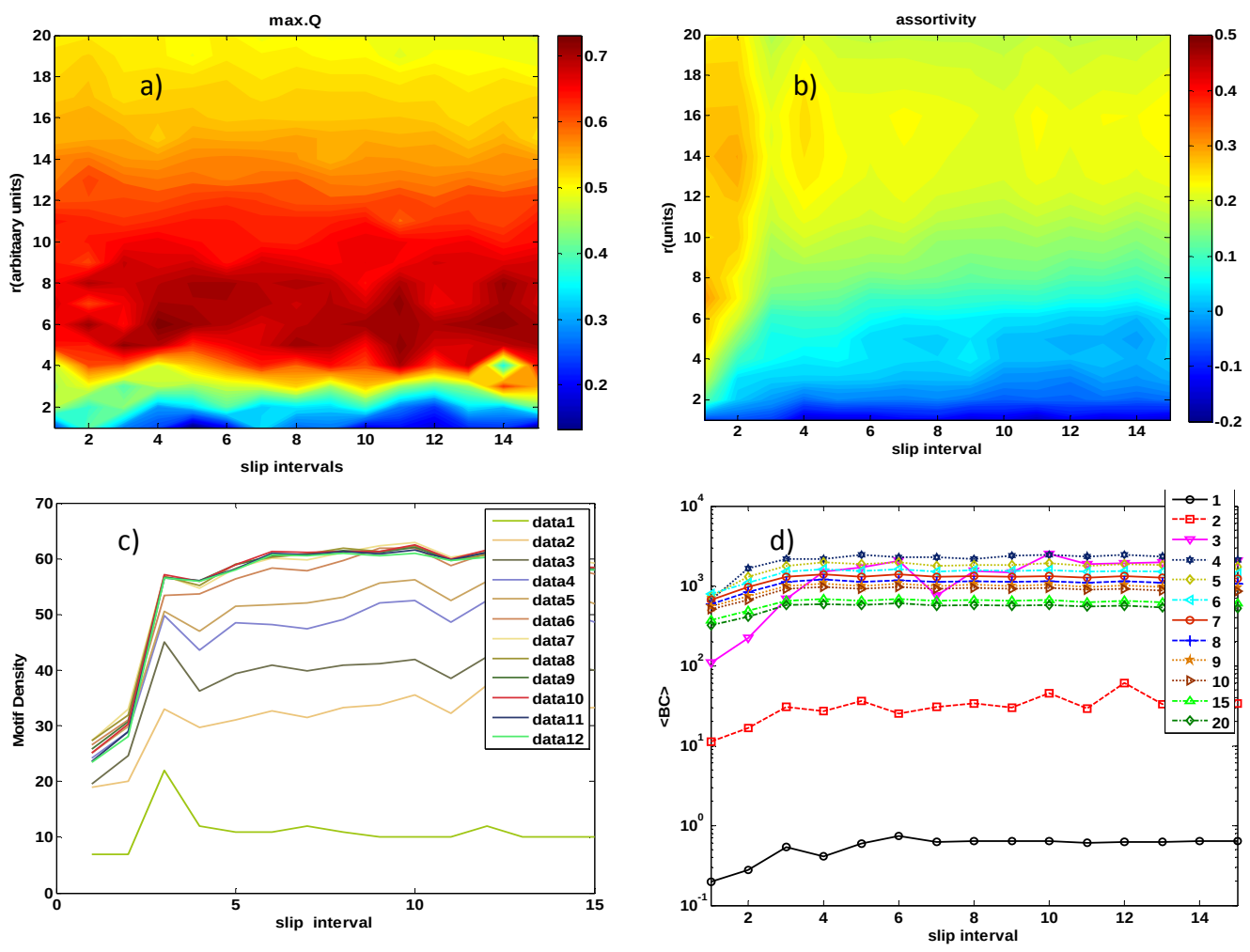

Fig 8. a) maximum modularity variation in influence radius and slip intervals; b) assortativity landscape :transition from correlated networks to uncorrelated pair-wise networks occurs around r~6 (arbitrary units) ;c) the density of 3 points-motifs versus slip interval (~displacement)and for different effective radius (data1 to data 12 correspond to r=1 to 20;see(d) legend) ;d) the average of betweeness centrality versus displacement .

## 6. CONCLUSIONS

As a conclusion to our study, we introduced aperture-friction networks over aperture patterns. We introduced new universalities with respect to the evolution of dry frictional interfaces: the scaling of local and global characteristics and gamma distribution of critical-curvature points on shear displacement field. Our results showed how the relatively highly correlated “elements” of an interface can reveal more features of the underlying dynamics. We presented a self-organized model to represent the evolution of the unraveled networks; standing on Turing patterns. As another result to our study, the interactions of sources were mapped in graphs where we found transition to slip-weakening was scaled with maximum variation of motifs density. This showed the special points undergo pair-wise interactions, changing the topology of the fracture energy landscape.

## ACKNOWLEDGMENT

We would like to acknowledge and thank Prof. M. Sharifzadeh (Amirkabir University of Technology, Tehran) who provided the data set employed in this study.

## REFERENCES

[1] Bowden, F. and Tabor, D.: The Friction and Lubrication of Solids, Oxford University Press, New York, 2001.

[2] Dieterich, J.: Time-dependent friction and the mechanics of stick slip, Pure and Applied Geophys. 116, 790–806 ,1978.

[3] Rubinstein, S. Cohen, G. and Fineberg, J.: Detachment fronts and the onset of friction, Nature. 430, 1005-1009, 2004.

[4] Dieterich, J. H., Kilgore, B. D.: Direct Observation of Frictional Contacts - New Insights for State-Dependent Properties, Pure Appl. Geophys. 143, 283-302, 1994.

[5] Ghaffari, H.O. Sharifzadeh, M. Evgin,E. and Fall, M.: Complex networks on a Rock joint, ROCKENG09: Proceedings of the 3rd CANUS Rock Mechanics Symposium, Toronto, May 2009 (Ed: M.Diederichs and G. Grasselli),Toronto,Canada,2009.

[6] Ben-David, O. and Fineberg, J.: Short-time dynamics of frictional strength in dry friction, Tribology Letters. 39, 235-245, 2010.

[7] Budakian, B. and Putterman, S. J.: Correlation between charge transfer and stick-slip friction at a metal Insulator interface, Phy.Rev.Lett.85, 1000, 2000.

[8] Filippov, A.E. Klafter,J. and Urbakh,M.: Friction through dynamical formation and rupture of molecular bonds, Phy.Rev.Lett.92, 135503,2004.

[9] Rubinstein, S. M. Cohen, G. and Fineberg, J.: Visualizing stick-slip: experimental observations of processes governing the nucleation of frictional sliding, J. Phys. D: Appl. Phys. 42 214016, 2009.

[10] Ghaffari, H.O. Nabovati,A. Sharifzadeh,M. and Young,R.P.: Fluid flow analysis in a rough fracture (type II) using complex networks and lattice Boltzmann method, PanAm-CGS 2011, Toronto, Canada, 2011.

[11] Ghaffari, H.O. Sharifzadeh, M. and fall, M.: Analysis of aperture evolution in a rock joint using a complex network approach , Int J Rock Mech Min Sci.47, 17,2010.

[12] Ghaffari,H.O. Thompson,B.D. and Young,R.P: Frictional interfaces and complex networks, AGU Fall Meeting 2011, http://eposters.agu.org/files/2011/12/2011-AGU-poster-Compatibility-Mode.pdf

[13] Chikkadi,V. Wegdam,G. Bonn, D. Nienhuis, B. Schall P.: Long-range strain correlations in sheared colloidal glasses, Phy.Rev.Lett.107 ,19, 198303,2011.

[14] Maloney, C.E. and Robbins,M.O.: Long-ranged anisotropic strain correlations in sheared amorphous solids Phy.Rev.Lett.102, 225502,2009.

[15] Ghaffari,H.O. and Young,R.P: Network–Configurations of Dynamic Friction Patterns, http://arxiv.org/abs/1201.3136 2012.

[16] Drazin, P. G.; Johnson, R. S. Solitons: an introduction (2nd ed.). Cambridge University Press,1989.

[17] D. Bonamy , Intermittency and roughnening in the failure of brittle heterogeneous materials Intermittency and roughnening in the failure of brittle heterogeneous materials Journal of Physics D: Applied Physics 42, 2114014 (2009).

[18] Johnson, K.L., Contact Mechanics, Cambridge University Press, Cambridge, 1987

[19] Sharifzadeh, M. Mitani, Y. and Esaki,T.: Rock joint surfaces measurement and analysis of aperture distribution under different normal and shear loading using GIS ,Rock Mechanics and Rock Engineering. 41, 229 , 2008.

[20] Sharifzadeh, M.: Experimental and theoretical research on hydro-mechanical coupling properties of rock joint, PhD thesis, Kyushu University, Japan, 2005.

[21] Ghaffari, H.O., and Young, R.P., Topological Complexity of Frictional Interfaces: Friction Networks, http://arxiv.org/abs/1105.4265 (2011) ; Z. K.Gao, and N. D.Jin, Nonlinear dynamic analysis of large diameter inclined oil-water two phase flow pattern, Phys. Rev. E 79, 066303 (2009).

[22] Guimerà R., & Amaral L. A. N., Functional cartography of complex metabolic networks, Nature 433, 895–900 ,2005.

[23] Newman M. E. J., Modularity and community structure in networks, Proc. Natl. Acad. Sci., 1038577–8582, 2006.

[24] Turing, A. M. The chemical basis of morphogenesis. Phil. Trans. R. Soc. Lond. B 237, 37–72 (1952) ;Mortazavi, V.; Nosonovsky, M. Friction-induced pattern formation and Turing systems, Langmuir, 27, 4772– 477,2011.

[25] Zapperi S., Vespignani A., and Stanley H. E., Plasticity and avalanche behavior in microfracturing phenomena Nature (London) 388, 658, 1997.

[26] Andresen, C. A., Hansen, A. & Schmittbuhl, J. Ridge network in crumpled paper. Phys. Rev. E 76, 026108 (2007);